\documentstyle[12pt]{article}
\begin{document}
\title{ETC With a GIM Mechanism}
\renewcommand{\baselinestretch}{1.0}
\author{Lisa Randall
\thanks{This work is supported in part by funds provided by the U.S.

Department of Energy (DOE) under contract \#DE-AC02-76ER03069
and in part by the Texas National Research Laboratory
Commission under grant \#RGFY92C6.\hfill\break
National

Science Foundation Young Investigator Award.\hfill\break
Alred P.~Sloan

Foundation Research Fellowship.\hfill\break
Department of Energy Outstanding Junior

Investigator Award.\hfill\break
CTP\#2112\hfill September 1992}\\
Massachusetts Institute of Technology\\
Cambridge, MA 02139\\
}
\date{}
\maketitle
\vskip-5in
May 1992 \hfill  MIT-CTP\#2112
\vskip5in
\renewcommand{\baselinestretch}{1.2}
\abstract{We construct an extended technicolor model of quarks and
leptons
which preserves a GIM mechanism. Furthermore,
there is only a minimal technicolor sector, in accordance with recent
precision measurements of

electroweak parameters.

We also show how to incorporate custodial SU(2) and  massless
neutrinos
into such a model.}
\thispagestyle{empty}
\newpage

\newcommand{\hc}{{\rm h.c.}}
\newcommand{\gev}{{\rm GeV}}
\newcommand{\tr}{\mathop{\rm tr}\nolimits}
\newcommand{\uone}{\mbox{U(1)}}
\newcommand{\su}[1]{\mbox{SU(#1)}}
\newcommand{\cf}{{\it c.f.}}
\newcommand{\half}{{\textstyle{1\over2}}}

\newcommand{\third}{{\textstyle{1\over3}}}

\newcommand{\fourth}{{\textstyle{1\over4}}}

\newcommand{\bra}[1]{\left\langle #1\right|}
\newcommand{\ket}[1]{\left| #1\right\rangle}
\newcommand{\vev}[1]{\left\langle #1\right\rangle}
\newcommand{\hatk}{\hat{k}}
\newcommand{\hatq}{\hat{q}}
\newcommand{\hatl}{\hat{l}}
\newcommand{\imn}{{I_{\mu \nu}}}
\newcommand{\beq}{\begin{equation}}
\newcommand{\eeq}{\end{equation}}
\def\barr{\begin{eqnarray}}
\def\ear{\end{eqnarray}}
\newcommand{\ba}{\begin{array}}
\newcommand{\ea}{\end{array}}
\section{Introduction}
It has been more than fifteen years
 since the original introduction of the technicolor (TC)
idea \cite{TC}. It was immediately recognized however that although
technicolor successfully breaks electroweak symmetry, the generation
of

fermion masses require extensions of technicolor models
which incorporate flavor dynamics \cite{ETC}. Since that time, many
proposed mechanisms for flavor dynamics have been invented;

however it is not clear that we have made progress towards
understanding the strange pattern of masses and mixing angles.
The only thing which  is clear is that there are very strong
constraints
on technicolor models. We know that I) flavor dynamics
is hidden from the low energy world, except as manifested
in the quark and lepton mass matrices \cite{ETC,wellknown} and II)
anything
more than a minimal technicolor sector is unacceptable
\cite{precise}.
These facts are determined experimentally. We conclude that
one of two things must be true; either technicolor is
not a viable alternative to the standard model or current
technicolor ideas are too limited.  If one is
to not to abandon the technicolor idea  as a viable alternative to
the
standard model, it is necessary to demonstrate the
possibility of accommodating these constraints in a realistic theory.
If this is indeed possible, and there is a class of technicolor
theories which are sufficiently different to successfully
evade current constraints,
the arguments and prejudice against
technicolor theories based on recent precision electroweak tests
would not apply.

In this paper, we present  extended technicolor models which
not only have a minimal technicolor sector but automatically
incorporate a GIM \cite{gim}  mechanism.

 The idea is simply that there
are separate gauged
Extended Technicolor  (ETC) groups for the left handed fermions,
the right handed up type, and the right handed down type fermions.
Such an idea for maintaining a GIM mechanism was first presented
in ref. \cite{technigim}.
The separate ETC gauge symmetries are broken so that global
symmetries

 are maintained.
 The global symmetries are
broken only by the terms which are necessary to generate
quark and lepton masses   and by the gauging of the electroweak,
strong, and technicolor gauge groups.

The class of models just described includes as a subset the
Composite Technicolor Standard Model (CTSM)

models proposed by Chivukula and Georgi in ref. \cite{ctsm1}.

We will however refer to this class of models as ETSM (Extended
Technicolor Standard Model) to emphasize that fermions need not
be composite, and moreover, the assumption of the full global
symmetry group might not be an essential feature of the model,
as it is not mandated by experiment.  In this paper, we construct
a model of quarks and leptons which may be interpreted as either
a CTSM or ETSM model.  It includes both quarks and leptons.  As
indicated above, there are separate ETC groups but a single
technicolor
doublet.  Although the model is cumbersome, in that it incorporates
many gauge groups, it does have some nice features.  In particular,
the up and down sectors can be treated symmetrically, leading to
a very simple embedding of the gauged U(1) which is necessary to
generate hypercharge.

To be specific,
we do not explicitly construct the flavor symmetry breaking
physics. What we do is show that it is possible for
flavor physics to produce quark masses and mixing angles
without generating other large flavor violation within

the context of an extended technicolor
scenario.
This gives rise to interesting model building questions
on how to generate flavor physics,

but flavor changing neutral currents are not problematic so
long as the flavor physics scale is sufficiently high.

We illustrate these ideas with two models.
The first incorporates only QCD like dynamics, so

vacuum alignment and anomaly cancellation are clear.
Essentially the simplest model one can write down based
on the above assumptions works.
However, the scale of flavor physics turns out to be quite low.

In the second model, the flavor physics scale is much higher.
However, this model requires nontrivial
assumptions about vacuum alignment.

Our models should have a striking  experimental signature,
namely the presence of additional light (less than about a TeV)
fermions and/or scalars. The precise spectrum is model dependent,
but could be visible in current experiments.
We also briefly outline some of the new model building ideas
which can be incorporated in this class of models. In particular,
we will show how to construct a model with deviation in the $\rho$
parameter from unity no greater than in the standard model. We
also show how to keep neutrinos massless with a see--saw like
mechanism.
Finally, we

suggest how the scale of leptons and quarks, or of generations,
could be distinguished.

In the next section we describe the general mechanism,
following which we present a simple model.
The following section presents
a more complex model.
The fifth section looks at the additional gauged U(1). The sixth
section looks at some of the phenomenological consequences of
this model, in particular the light additional fermions and
pseudogoldstone
bosons which are present.
In Section 7, we discuss neutrino masses and the $\rho$ parameter.
The following section suggests the richness of this class of models,
in which novel ideas for distinguishing generations can be
incorporated.
We conclude in the final section.

\section{The Mechanism}

Let us briefly recall why mass generation through extended
technicolor interactions generates excessive flavor changing neutral
currents.  In order to obtain different masses for different
generations, the ETC gauge bosons whose
exchange yields the four fermion

operators (two technifermions and two light
fermions) responsible for light fermion masses
must be  nondegenerate.  That
is, fermion masses are proportional to the inverse square of the ETC
gauge boson mass. The hierarchy of quark and lepton masses
requires a hierarchy of ETC gauge boson masses.  But if the ETC
gauge boson masses are different, a unitary rotation to go from a
weak
to a mass basis is a nontrivial operation; flavor changing effects
do not cancel between different generations.  The only way to
eliminate these flavor changing effects is for the ETC gauge
bosons to be nearly degenerate in mass. But in this case, fermions
would also  be degenerate, unless ETC exchange did not
directly generate the four fermion operator which gives mass.

The way to achieve the latter goal is obvious. The ETC generators
which generate a rotation between the left handed quarks [leptons]
and
techniquarks must be different than those which generate rotations
between the right handed quarks [leptons] and techniquarks.  It
is then possible that the ETC gauge bosons are approximately
degenerate,
but that the mass mixing between left and right handed gauge bosons
is different for the different generations, thereby generating a mass

hierarchy.  It is then possible to generate a

mass hierarchy while suppressing flavor changing effects through the
near
degeneracy of the ETC gauge bosons.

This idea was also  incorporated
in the models of  ref.
\cite{technigim,us,japan}.

Let us enumerate the ingredients which should be required
 of such a model. We first
discuss these ideas  generally; they are illustrated in subsequent
sections.
First, we must have separate ETC gauge groups. We will call
these $ETC_L$, $ETC_U$, and  $ETC_D$.  Because we want to
have only a single technicolor representation, we want electroweak
SU(2) to commute with the ETC groups.  Furthermore, these
groups must give mass to both quarks and leptons.  In a model
with right handed neutrinos, the ETC groups will be SU(12+n)
for the $ETC_L$, $ETC_U$, and $ETC_D$ groups.
Here 12 includes three generations of quarks and leptons.
The technicolor
gauge group is SU(n).

Second, the ETC gauge groups must be broken.  From the discussion
above, we know that if a GIM mechanism is to hold, the ETC
gauge groups must break in such a way that the gauge generators
are nearly degenerate. That is, we want three SU(12)

{\it global} symmetries to survive.   This is of
course stronger than what is really mandated by flavor changing
phenomenology; for simplicity, we assume the full global
symmetry groups for now.  We also need the
separate SU(n) gauge groups embedded in the three ETC gauge groups
to break to a common technicolor.
In the models we construct, we do not permit elementary scalar
fields.  Therefore, we assume the ETC gauge groups break due
to fermion chiral condensation.  Our models will therefore include
strong gauge groups, which we call $SU(S_U)$, $SU(S_D)$, and
$SU(2S_L)$
which induce chiral symmetry breaking of the ETC groups.
These groups get strong at  scales, $f_L$,

$f_U$, and $f_D$ of approximately 1 TeV.

It is possible to combine the strong groups in variants of
our model; for example the $S_U$ and $S_D$ gauge groups could be the
same.
The symmetry breaking structure is such that a global $SU(12)_L$
combines with the $SU(12)$ subgroup of the $SU(12+n)_{ETC}$ group
to leave an unbroken $SU(12)$ global symmetry which maintains
the near degeneracy of the $ETC_L$ gauge bosons. This structure
is replicated for the up and down ETC gauge groups. Furthermore,
there is a gauged $SU(n)$ group which combines with the three
$SU(n)$ subgroups of the three ETC groups; that is, there is an
unbroken gauged
SU(n) whose generators are a linear combination
of the SU(n) generators of the ETC groups and the separate SU(n).
The SU(12) global symmetries are a feature of the simplest models
of this sort; it is only necessary that some subgroup of the global
group remain as an approximate symmetry to protect against flavor
changing neutral currents.

The TeV scale $f$ sets the mass
scale for the ETC gauge bosons.

The final requirement in order to generate masses is that
the global SU(12) symmetries are only approximate.
They must be broken in such a way that the $ETC_L$ generators
can mix with the $ETC_U$ and $ETC_D$ generators.  In the simplest
models, the mixing will be proportional to quark and lepton
masses, so the mass hierarchy is a direct consequence of the
hierarchy
of mixing.  More elaborate models would incorporate different
symmetry breaking scales; that is the full global symmetry group
is not necessary.

Although the mass mixing is a consequence of multifermion operators
generated at high energy,
the only large flavor violation which filters down to the
low energy world will be mass terms.  Other operators will be
suppressed either by a large flavor physics scale or by
light fermion masses.
Therefore, the standard GIM mechanism is maintained.
Large flavor violations will occur only because one cannot
simultaneously diagonalize the up and down quark mass matrices; this
means flavor violation will be suppressed,
similar to the standard model.
Of course, if we do not maintain the full global symmetries, flavor
changing effects could be larger.

 The net result of ETC exchange
and mixing between left and right handed ETC gauge bosons
is a four fermion operator of the form
\beq \label{fermionmass}
{1 \over 4 \pi  f^3} \overline{q} M q \overline{T} T
\eeq
where $M$ is a symmetry breaking parameter proportional to the quark
mass,
$q$ are light fermions,  $T$ are technifermions, and $f$ is
determined by the ETC symmetry breaking scale.  This yields a fermion
mass

\beq
m \approx M \left({v  \over f}\right)^3 .
\eeq
 Here we have simplified the
estimate by assuming a single scale $f$.

So to summarize, we assume independent
 ETC groups on the left and right handed fermions.
 In the simplest models, these gauge bosons mix
 proportionally to mass. Such a mechanism is
  incorporated in ref. \cite{ctsm1}. Here we incorporate
this mechanism into a more general context.  We take
a strict effective field theory point of view. The phenomenological
constraints from FCNC and precision analysis of electroweak
parameters tell us to separate the ETC groups.

We then build in the remaining elements which are necessary
to construct a consistent model.
  In the
following section, we restrict our analysis by focusing
on two models which demonstrate the feasibility of this mechanism.

\section{A Model}
To describe the model,
we employ the  so called ``moose"
notation. This notation is a simple
way to describe the symmetry groups (local and global),
the fermion content, and anomaly cancellation of a model
constructed from fundamental and antifundamental representations
of gauge groups.
In particular, every circle  is
a gauge symmetry group and the line segments emanating from this
circle represent a fermion transforming as a
fundamental[antifundamental]
representation according to the orientation of the arrow on the line.
Anomalies are cancelled if the same number of fermions enter and
leave
the circle representing the gauge symmetry group.
This is illustrated in Figure 1.

The cumbersome appearance of
models constructed in this way is inseparable from the fact that
one has simplified the analysis of models by separating them
into pieces which can be analyzed independently. So although
the interpretation of the models can be quite simple,

they comprise many components in general.  However, the
interpretation
at any given scale is often straightforward.

But even with

an effective field theory viewpoint,

it is still necessary to prove the
existence of consistent models.  In ref.

\cite{us,japan}, difficulties in constructing CTSM models were
discussed. In this section and the subsequent one, we show
that one can resolve the problem of anomaly cancellation
which prevented the successful construction of models with both
quarks and leptons.  Although the models we construct can
be interpreted as composite models exemplifying the CTSM ideas,
they have a more general interpretation simply as ETC models with
a GIM mechanism.

We now present the model.  This is given in Figure 2.

We first enumerate the gauge groups. They are ordered from top to
bottom according to the scale at which it is assumed strong groups
break.
 Starting from the bottom
of the ``moose", there are two $S-1$ gauge groups.  These groups
are not associated with flavor generation, but appear to
be necessary for a phenomenology consistent
with anomaly cancellation. We discuss this further later.

There is a weakly gauged SU(2) group. This is the standard
electroweak SU(2).  Notice that it is factored out of ETC,
so only one fermion line carries electroweak SU(2).

There are three ETC gauge groups. These are the three
SU(12+n) groups labeled $ETC_U$,
$ETC_D$, and $ETC_L$.  There are three gauged groups labeled
$S_U$, $2S_L$, and $S_D$.  These groups get strong at the scales
$f_U$,
$f_L$, and $f_D$

of approximately 1 TeV.  There is a gauged SU(n) group
(this n is the same as the n of 12+n in the ETC group). This
group is responsible for breaking the three separate SU(n) subgroups
of the ETC gauge groups to a common technicolor gauge group; that
is a linear combination of this SU(n) and the SU(n)'s embedded in
the extended
 technicolor groups is technicolor. Finally, color and hypercharge
 are weakly gauged. We discuss hypercharge in Section 5.

Now we discuss the fermions.  Again we start from the bottom of the
moose.

There are  fermions which transform as  $(\overline{n+12}_U, S-1_U)$
and $(\overline{n+12}_D, S-1_D)$
and

fermions which transform as an $(\overline{S-1}_U, n+12_L)$

and an $(\overline{S-1}_D, n+12_L)$, where
the $n+12$'s refer to different
ETC groups.  These fermions are necessary for anomaly cancellation

of the ETC groups. In this model,
they are the light fermions whose
exchange mixes the left and right handed ETC gauge generators
so that physical fermions can acquire a mass.

The physical fermions (that is light quarks, leptons, and
technifermions)
are the  $(\overline{n+12}_U)$ (three up quarks), the
$(\overline{n+12}_D)$ (three
down quarks and three leptons), and the $(2,n+12_L)$ (three quark
electroweak
doublets and three lepton electroweak doublets).

There are $(\overline{S_U}, n+12_U)$, $(\overline{S_D}, n+12_D)$ and

$(\overline{n+12}_L,2S_L$) fermions.

These will condense with the fermions which carry global
flavor symmetry when the two SU(S) and the SU(2S) groups get strong.

There are $(\overline{12_U}, S_U)$,

$(\overline{12_D}, S_D)$, and $(2S_L, 12_L)$
fermions.  Here, $12_U$, $12_D$, and $12_L$ are global flavor
symmetry groups.
When the $S$ and $2S$ groups
get strong, the degeneracy of the ETC gauge boson masses will
reflect this global symmetry.  In this model, the global symmetry
is weakly broken by six fermion operators, involving the
$(\overline{12_U},S_U)$,
$(\overline{S_U}, n+12_U)$, $(\overline{n+12}_U, S-1)$,
$(\overline{S-1},
n+12_L)$,
$(\overline{n+12}_L, 2S_L)$, $(\overline{2S_L}, 12_L)$ fermions
 and similarly
 with $(\overline{12_D}, S_D)$. This is the weakest feature of the

``model". We assume

the existence of these operators, but do not address the
question of  their origin.

There are also fermions which transform as $(\overline{n}, S_U)$,

$(\overline{n}, S_D)$, and $(\overline{2S_L}, n)$.  Because of the
gauged SU(n),
when the ETC groups break, technicolor will remain as  a common
gauged SU(n) group.  The single SU(n) group allows an automatic
approximate custodial SU(2) symmetry.

In this  model, there is a scale $f$ where the ETC gauge symmetries
break (in principle there can be three independent scales). This
is the scale where the $S$ and $2S$ groups get strong.

Because all we have is QCD dynamics, we can analyze how things work.
After
the ETC gauge symmetries are broken, the ETC generators are
approximately
degenerate and there
is a single gauged technicolor SU(n).
  The light fermions remaining in the theory are the
standard model fermions and technifermions, plus the fermions which
transform under the $S-1$ gauge group.  The fermions which

felt the SU(S) and SU(2S) gauge interactions are bound into heavy
fermions,
which we can ignore.

Because the fermions transforming under the strong groups
condense,
the residue of the six fermion operators are mass terms for the
fermions
which carry the $S-1$ gauge symmetry.

These take the form

\beq
\overline{(\overline{n+12}_U, S-1_U)} M (\overline{S-1}_U, {n+12}_L).
\eeq
These
mass terms are what allow the left and right handed ETC generators to
mix.
The fact that the gauge symmetries break not to exact,
but to approximate global symmetries,  is the key
to mass generation.

This is similar to  the  CTSM scenario.

A fermion gets

a mass as follows. $ETC_L$ exchange generates a four
fermion operator with a light fermion, a technifermion, and
two of the non standard model fermions which transform under
the S-1 group.
Because the fermions transforming under the S--1 group are massive,
the ETC generators of the left and right handed
fermions
can mix.  A four fermion operator
involving quarks or leptons and technifermions is
thereby generated. When technicolor breaks, the quarks and leptons
become massive.

Notice that the fermions are not composite in this model. The
 independent ETC gauge generators of the left
and right handed symmetry groups mix, but only through terms
proportional
to quark and lepton masses.  The maximum value for the mass
is then $M v^3/f^3$ where $M$
is the mass for the internal fermion line,
$v$ is the technicolor scale, and $f$ is
the scale at which the $S$ and $2S$ interactions get strong. Here we
have assumed they are the same; if not, there could be additional
suppression
factors.

We now consider constraints on the model.
We have already emphasized
that precision tests of the electroweak
sector are not a problem, since we have only
a single technicolor representation. The possible exception
would be if the ETC scale were so low as to induce large corrections
to the $\rho$ parameter. We discuss this further in Section 6.

Another constraint of course comes from the masses themselves,
which determine the possible scales for the dynamics. Unfortunately,
because the top quark mass is so large, the flavor and strong scales
cannot be too much larger than the electroweak scale, which in this
model
implies that
the scale of flavor physics
cannot be pushed up too high.  From eq. (2),
the top mass cannot be much bigger than $4 \pi v^3/f^2$, where
$v$ is the electroweak symmetry breaking scale and $f$ is the ETC
breaking scale.  Because other interactions mediated by ETC gauge
bosons constrain $f$ to be greater than about 1 TeV, this
would be the approximate scale of symmetry breaking. This means
the scale of the multifermion operators cannot be much bigger
or the top quark mass would be too small.

Because there
is not a clean separation of flavor scales and ETC scales
in this model, the flavor physics should be incorporated if
one is to really view it as a model.  Here, we see this is a
nontrivial
problem, because the flavor structure is very constrained due
to its low scale. In particular, large flavor changing effects
could be induced if in addition to the six fermion operators
already discussed there exist four fermion operators involving
only the $12_U$, $12_D$, or $12_L$ fermions.

Because the flavor physics
is given at a low scale, to specify the theory would really
require  incorporating it.

In the next section, we present a model with a flavor scale
which is several orders of magnitude greater than the ETC
scale. In that model, the problems just discussed do not arise.
Flavor changing effects are suppressed either by the high
flavor scale or by quark mass.
Nevertheless, the example in this section works well

as a simple illustration that ETC with a GIM mechanism is possible.

The final constraint on the model
comes from the
  additional
light fermions in the theory,
which seem necessary for a consistent anomaly free
theory.

We assume the $S-1$ group
also gets strong, so the
light fermions are replaced by light pseudogoldstone
bosons. We discuss the phenomenology of the pseudogoldstone bosons
in Section 4.

So to summarize what we have learned, we have shown that it is
possible to construct a model in
which flavor physics is
parameterized in
mass terms. We have successfully generated a model with
both quarks and leptons.
Notice the virtue of this model.
All the flavor physics has been completely isolated from the
sector which carries electroweak SU(2).  The only fermion
doublets are the usual fermions and one family of technifermions. To
 construct a model with quarks
and leptons does
nevertheless require the presence of additional fermions transforming
under the ETC groups, in order to ensure anomaly cancellation. There
could be a very interesting, but model dependent scalar spectrum from
the pseudogoldstone bosons generated at this scale.
This  requirement should  yield testable experimental
consequences, which we discuss in Section 4.

\section{Another Model}

In this section, we construct a model with a high flavor scale.
It is very simple to construct  models based on the ideas
presented in refs. \cite{us,japan}.  As was emphasized in ref.
\cite{us},
the generation of  a nontrivial KM matrix requires that the
independent
global symmetries of the left and right handed fermions be broken
not by mass terms, but by higher dimension ({\it e.g.} four fermion)
operators.  This is because if there are only mass terms, flavor
mixing can be diagonalized away. Models which incorporated
four fermion operators were given in ref. \cite{us,japan}.  However,
these models only included quarks. Here, we show that it is simple
to extend the models to include both leptons and quarks. The idea
again is that the cancellation of anomalies requires the presence
of additional fermions transforming under the three ETC groups.

Since much of the analysis of this model is similar to what was
described in ref. \cite{us,japan}, we will not present excessive
detail. The reader should refer to these papers for a more
comprehensive
analysis.
Because the model of ref. \cite{japan} is simpler, we present only
one
example based on
this model. Notice that

one can readily construct variants on the model depending on whether
one
assumes independent SU(S) groups.

The model is based on the
same moose as the previous section, given in Figure 2.

Again, the real fermions and technifermions are the single lines
at the bottom of the diagram; that is they transform under the
SU(n+12)
gauge groups labeled $ETC_U$, $ETC_D$, and $ETC_L$.

The difference in the model is in the structure not visible in the
moose,
namely the structure of the flavor symmetry breaking multifermion
operators.
As opposed to the model of the previous section, we assume
the flavor physics has its origin at a high scale, $\mu_F$,
which can be several orders of magnitude larger than the
ETC scale \cite{us}.

We assume the global flavor symmetries are
broken by four fermion operators.

These four fermion operators take the form
\beq
{1 \over \mu_f^2} \tr \left( \lambda_U(12_L,\overline{2S_L}) C
\gamma^0
(S_U, \overline{12_U})\right)
\left( (S_U, \overline{n}) C \gamma^0 (n, \overline{2S_L})
\right)^{\dagger}
\eeq
and similarly for down.

We assume the same symmetry breaking pattern as the model of ref.

\cite{japan}.      The
SU(n) group gets strong first at a scale $\mu_n$.
This
causes the  $SU(2S_L)$ group

to break. There are heavy gauge bosons and fermions associated
with this scale. The single $SU(12_L)$ splits into two
approximate global SU(12) groups.
However, the approximate symmetry
is broken by four fermion operators generated

by exchange of SU(2S) gauge generators.  These take the form
\beq
{1 \over \mu_S^2}\tr
\left(\overline{12_{LU},\overline{SU})}\gamma^\mu
(12_{LD},\overline{SD})\right)

\tr\left(\overline{(12_{LD},\overline{S D})}\gamma_\mu (12_{LU},
\overline{S U}\right).
\eeq
These are important, for they are how the theory \lq \lq remembers"
that
there were not originally two independent global SU(12) groups.

This is of course the same structure as in ref. \cite{japan}. If they
were not there, the KM matrix would be trivial.
Once the SU(n) group becomes strong,
the theory looks like there are two mass terms in
the low energy theory below the scale $\mu_n$, one for the up
type flavor fermions  and
one for the down type.  These mass terms take the form
\beq
{4 \pi \mu_S^3 \over \mu_F^2} \tr\left(\lambda_U(12,\overline{S_U}) C
\gamma^0

(S_U,\overline{12})\right).
\eeq

Now, when the $S_U$ and $S_D$  groups get strong, it is assumed that
the ETC gauge groups are broken to global groups, again
by chiral condensation.  In this model, this
is a nontrivial assumption about the vacuum alignment. Also,
when the two $12_L$
groups break the single ETC group on the left handed fermions,
there is an additional alignment question. This however is what is
required
in order to generate

the possibility of a nontrivial KM matrix. A similar  vacuum
alignment
was studied in ref. \cite{us}, where it was shown that
the competition between the two types of four fermion operators
naturally leads to  a KM matrix of the right sort, that
is where the families are correlated in mass, and the KM angles
are bigger for the lighter generations.  The vacuum alignment in this
model merits further investigation.

The final stage of symmetry breaking is the condensation of the S--1
group, to eliminate light unobserved fermions.

Note that despite the complicated appearance of this model, there
are in fact very simple features. A detailed understanding
of these features indicates the directions for future model
building. We outline what we consider the important aspects
of this model now, and suggest

variations in Section 8.

There are several scales involved in this type of model.  We
only begin to describe the model below a flavor physics scale,
$\mu_F$.
In the models as they presently stand, this scale is restricted
to be less than approximately $1 ({\rm TeV}/f)^5 (100 \gev/m_t)^{3/2}
4000

{\rm TeV}$
\cite{us}.
Eventually one would like to understand the origin of four fermion
operators at this scale, but for now, they are assumed.

There is  a scale $\mu_n$ which  which sets the scale
for the operators which are necessary for there

to exist a nontrivial KM matrix.

Recall that multifermion operators permit the existence
of a nontrivial KM matrix.

As we see by comparing this model to that of the previous section, if

the multifermion operators become mass terms at a high scale,
then such a splitting scale is probably required.
In the first model, where the flavor physics scale is low, this

additional scale was not necessary.
However, in the model with a high scale of flavor physics,
mass terms are the  operators
of lowest dimension and can transmit

flavor violation to low energies. But mass terms
alone are not sufficient, since the KM
matrix would then be trivial \cite{us}.
The splitting scale, if necessary, should be less
 than a few hundred GeV \cite{us}.

There is a scale of order a TeV where the ETC groups break. At
this scale the global symmetry breaking is transferred to mixing
between ETC generators, allowing the four fermion operators with
two light fermions and two
technifermions which yield light fermion masses.
If the model has an interpretation as a composite model, this would
be the fermion compositeness scale.

Finally, there is a low scale which could be of order 100 to 1000 GeV

where
light additional fermions necessary for anomaly cancellation are
removed
from the theory, being replaced by pseudogoldstone bosons.

We discuss this scale further in a subsequent section.  It is
interesting that in the models we have so far constructed, this scale
appears to be required.

Notice that this separation of the physics at the separate scales
makes it easier to interpret the model. Possible variations on this
most simple scheme are suggested in Section 8. However, even in
variations,
it is clear that one needs to input flavor physics in the form of
higher
dimension operators, transmit the flavor breaking to the low energy

world, and eliminate light unobserved fermions. We expect these
to be common features of all models of this sort.

\section{U(1)}

So far we have presented two illustrative models with an emphasis on
what we believe to be necessary in ETC models

of the ETSM type.
In subsequent

sections, we discuss features which could prove to be less universal.
These include the embedding of U(1) symmetries and the elimination
of light  nonstandard model  fermion representations of the ETC
groups.
In Section 6, we illustrate how one can naturally incorporate
massless
neutrinos. We also show how the $\rho$ parameter constraint
can be readily satisfied.

We first discuss U(1) symmetries.  In this model, a gauged U(1)
symmetry combines with a U(1) subgroup of the ETC group to become
hypercharge.
One interesting feature of the
model we have constructed is the simplicity of the embedding
of   the gauged U(1) symmetry.

In the model with both quarks and leptons,
the gauged U(1) charge  can be assigned independent of the size of
the
technicolor group. The charge of the physical left handed fermions
is 0, that of the up fields is $1/2$, and
that of the down fields is $-1/2$.  The remaining fermions
have charge $\pm 1/(2)$ or zero in accordance with anomaly
cancellation.
This seems a compelling feature of the models with both quarks and
leptons;
it should be contrasted to the U(1) charge assignments in the
models of ref. \cite{us,japan}, which contain only quarks and had
charges
which depend on $n$, the size of the technicolor group. Because the
charges are $\pm 1/2$ or $0$, anomaly cancellation is trivial.

The only additional nonanomalous U(1) symmetry acting on the physical
fermions is fermion number; that is, the symmetry which counts
quark plus lepton plus techniquark number.  There is no axion in
this simple model.

\section{Light Fermions and Pseudogoldstone Bosons}

 We now consider the phenomenology associated with the fermions
transforming
 under the S--1 gauge groups.  It is readily seen that  if the
 S--1 group did not get strong,
some of these fermions would be too light, since their mass is
directly related to the mass of the standard fermions.
However if the S--1 groups get strong, fermions lighter than
the S--1 scale
 are eliminated in favor of pseudogoldstone bosons.

In this model, there is both an upper and a lower bound on the
scale at which the S--1 groups get strong.  The lower bound comes
from the nonobservation of the pseudogoldstone bosons.  The upper
bound comes from
 the fact that the mixing of the left and right handed ETC gauge
 bosons arising from the condensation of the S--1 group should
 not be greater than the mixing arising from the

flavor symmetry breaking

operators.  We assume there is some mechanism (see
the following section)  for preventing too large a neutrino
 mass so that the strongest constraint will be on the
  gauge boson
 whose exchange yields the electron mass.
 Let us estimate the contribution to the mass which comes
 from  S--1 condensation.  First note that
 ETC exchange yields four fermion operators with coefficient
 of order $1/f^2$.  When technicolor becomes strong, a four
 fermion operator involving two standard fermions and two
 fermions (left and right handed) transforming under the S--1 group
are induced.  Its coefficient should be of order $v^2/f^4$.
 When the S--1 group gets strong, the result is
 a contribution to the fermion mass of order $4 \pi \mu_{S-1}^3
v^2/f^4$,
where $\mu_{S-1}$ is the scale at which the S--1 group gets strong.

This yields the constraint
 \beq
m_e> {4 \pi \mu_{S-1} v^2 \over f^4}.
 \eeq
 This yields $\mu_{S-1} < 15 \gev (f/1.5 TeV)^{4/3}$.

However, this estimate relied on replacing three condensates
by $4 \pi \mu^3$, where $\mu$ is the scale appropriate to the
particular condensate, so it is very likely this estimate can
be off by an order of magnitude.

 Now let us consider the masses of the pseudogoldstone bosons in this
 scenario. We first
consider the mass of the neutral pseudogoldstone bosons
formed at this scale.

The lightest neutral pseudogoldstone boson

gets its mass from the flavor symmetry breaking associated with
the electron mass.
Its mass is approximately
$(m_e (f/v)^3\ 15 {\rm GeV})^{1/2}$,

 which is about 1 GeV
 and safe. The other neutral pseudogoldstone bosons

formed at this scale  are even heavier.

Notice that there are additional pseudogoldstone bosons formed
at the technicolor scale.  These get mass from the SU(S--1), SU(2)
and
U(1) gauge charges.  This constrains the S--1 charge at the
technicolor scale to be not too small.

 The charged pseudogoldstone bosons also carry color or technicolor
 charge, and therefore get a mass not only from global symmetry
breaking
 mass terms but also from symmetry breaking gauge couplings.

The pseudogoldstone bosons carrying technicolor are heavy. Those
 which are color triplets can probably have mass of order of the
symmetry
breaking scale,
 with the octets bigger by a factor of about 3.
There are reasonably severe constraints on colored charged
pseudogoldstone bosons and fermions.  Light leptoquarks
would be seen at $e^+ e^-$ colliders. Moreover, they could
be searched for in the same modes as have been used to
search for the top quark at CDF (although presumably only the like
sign
lepton searches would be relevant).  One therefore expects
the leptoquark pseudogoldstone bosons to be at least 100 GeV.
More detailed phenomenology will be considered in a future
publication.
If there are heavier leptoquarks, they might be found in the future
at the Tevatron or HERA.

 We conclude that it is likely that a symmetry breaking scale as low
as 15 GeV is very likely ruled out.  However, as we have
stated previously, the estimate of 15 GeV is not reliable to better
than an order of magnitude.  Furthermore, the low scale was
a result of the particular model which we presented. An important
question is how general is this prediction.
 We answer this by demonstrating the existence of a model
 in which the symmetry breaking scale could be larger.
 The model

is given in Figure 6, in which additional fermions transforming
 under the S--1 gauge groups, as well as a weakly gauged n and 12
groups

are included.
 It is assumed that when the S--1 groups get strong, the
nontechnicolored
fermions
 condense with the fermions transforming under the ETC groups and the
S--1

 groups.

The constraint on the symmetry breaking scale $\mu_{S-1}$ is now less
stringent, and arises only from consistency of the model. One
expects therefore that this scale is less than a TeV.  The colored
pseudogoldstone bosons will get a mass

squared of order $\alpha_1 \alpha_s \mu_{S-1}^2$, so are probably no
heavier than a few hundred GeV.

 So we conclude that there are certainly additional scalars
associated
 with the model, with mass between 1 GeV and 1 TeV. The precise
 spectrum must however be model dependent. We expect the experimental
 predictions should be considered in full generality if we are to
 test the idea, rather than a particular implementation.

 It should
 also be observed that there might be other models entirely, for
which there are only fermions in this mass range
 since the additional symmetry breaking was not an integral part of
the mass
 generation mechanism.
 However, we have not
 yet constructed such a model which is consistent with existing
phenomenology.

\section{More Model Building: Neutrinos and the Rho Parameter}

In this section, we discuss how to generate models with two
interesting features. First, we show how the model can
readily generate massless neutrinos.  Second, we discuss
the rho parameter constraint, and show how the model
can preserve to a large degree
custodial SU(2) even with flavor physics incorporated,
so that the deviations of the rho parameter from unity
is about the same size as in the standard model.

We first discuss neutrino masses.  There are essentially four ways
to get light or massless fermions.  First is to rely on the flavor
physics which produces the four fermion operators to generate small
global symmetry breaking terms for the neutrinos.  Second is
that the ETC group for the up type fermions is only SU(9+n) and
the S--1 anomaly is cancelled by  a fermion which does not
transform under the ETC group.  Third is to implement a standard
see--saw mechanism.  Fourth is to incorporate additional singlet
neutrinos in such a way that a U(1) symmetry to maintain
massless neutrinos is naturally preserved.

The first is of course possible.  However, it would then be
necessary to ensure that all terms which could generate neutrino
masses are small. In particular, this would severely constrain
the S--1 breaking scale.  The  second possibility works fine; however
 one then loses the nice embedding of hypercharge which is possible
when all the ETC groups are SU(12+n).  We have not yet discovered
how to incorporate a standard see--saw into these models, since
no four fermion operator with only right handed neutrinos can
be constructed which preserves gauge invariance.  We think
the fourth possibility, in which

there are exactly massless neutrinos,
might in fact be the nicest way
to maintain massless neutrinos.

In this model for neutrinos, one assumes the existence of three
additional singlet neutrinos.  If these exist, one can then
construct a gauge invariant four--fermion
operator suppressed by a scale $\mu_\nu$ from the singlet neutrino,
the two fermion representations of $SU(S_U)$, and the
right handed neutrino.  When the $SU(S_U)$ group gets strong, the
right handed neutrino and singlet neutrino get a Dirac mass.
The mass matrix in the neutral fermion sector then consists
of a massive Dirac fermion of which the left handed component
is a linear combination of the singlet fermion and the left handed
neutrino and the right handed component is the right handed neutrino.
The orthogonal linear combination of the singlet fermion and the
left handed neutrino is left massless.  The scale of the heavy
singlet fermion is naturally $4 \pi f^3/\mu_\nu^2$.  If it is
assumed that the mass term between the left and right handed
neutrinos
is comparable to that for the corresponding leptons, the flavor scale
for the new four fermion operators is no greater than
100 TeV.   The massless fermions are then primarily left handed
neutrinos. If there are no additional U(1) symmetries imposed, flavor
changing
effects would impose the strongest constraint on the scale $\mu_\nu$.
This will be considered in a future work.

We now consider the $\rho$ parameter constraint in these models.
Notice that a lower scale for the top quark mass is not excluded
experimentally \cite{randall}. One of the more stringent
constraints however could prove to be the $\rho$ parameter. In
our model we separated off the global SU(12)

groups from the SU(n), so that four fermion
operators involving the technifermion fields
were not necessary. This helps maintain an approximate custodial
SU(2) symmetry.

Technifermions
masses other than those arising from chiral condensation would
not automatically preserve a custodial SU(2) unless further
complications were introduced. However, even with the SU(n)
group separated off, violations of custodial SU(2) symmetry
can arise if $f_U \ne f_D$. Suppose they were extremely
nondegenerate,
for example, $f_U \ll f_D$. Then the operator
 generated by ETC exchange between right
handed $U$ quarks, namely
\beq
{1 \over f_U^2} (\overline{U}_R \gamma_{\mu} U_R)

(\overline{U}_R \gamma^\mu U_R)
\eeq
gives the $Z$ but not the $W$ a mass. Of course, there is
no way to evaluate this four fermion operator. If for the purpose of
approximation, we factorize the operator into the product of two

right handed currents, we get
\beq
\Delta M_Z^2={v^2 \over f_U^2} {e^2 v^2 \over 8 \sin^2 \theta_W

\cos^2 \theta_W}
\eeq
which implies
\beq
\delta \rho=1.6 \% ({1.5 {\rm TeV} \over f})^2
\eeq
 From Peskin and Takeuchi the 2$\sigma$ limit on $\Delta \rho$
\cite{pt2}
from the $W$ mass alone (assuming zero $S$) is 1\%.
We would conclude that either $f_U \approx f_D$, the top mass is
somehow enhanced so that $f_U$ can be higher, or we are close
to seeing deviations in the rho parameter. As emphasized in ref.
\cite{randall}, this is true for any model for which the top mass
is generated by ETC gauge boson exchange. The difference
here is the possibility that $f_U$ could be close to $f_D$, since
these scales do not directly determine the top and bottom quark mass,
thereby
suppressing the violation of custodial SU(2). Because the scales
do not directly give rise to fermion mass in this sort of model,
a near degeneracy of the $t$ and $b$ scales is of course permitted.
In fact, the $SU(S_U)$
and $SU(S_D)$
groups could be combined so that this is automatic. In this
case, violations of custodial SU(2) would be no larger than that
generated from the top quark itself.

We conclude that this category of models has the advantage
that it is easy to preserve custodial SU(2) symmetry while
allowing for very different top and bottom mass. This is unlike
ordinary technicolor, where strong interaction uncertainties could
suppress the rho parameter somewhat, but not to much less than 1\%.

 \section{Directions for Future Research}

 We observe that so far we have been very restrictive in our
 assumption about symmetry breaking scales, and in our assumption
 that the full global SU(12+n) symmetry groups be maintained.
 This is certainly a far stronger requirement than is mandated by
 phenomenological considerations.  Future
 research should exploit these possibilites. We suggest some ideas
 in this section.

 First, the scale at which the up and down SU(S)  groups get strong
need
 not be the same. This could for example naturally suppress down
 type masses if $f_D > f_U$.  Of course, in this case we might
expect we are close to seeing deviations in the $\rho$ parameter from
unity.

 Second, these type of models could readily incorporate notions of
``tumbling".
 For example, at the top of the diagram, there could be a chain of
 gauge groups which break so that the full SU(12) global
symmetry group is not preserved.
 However this would require flavor physics to produce multifermion
operators.
 A better possibility in this regard is that there is not necessarily
 a single $S_U$, $S_D$, and $S_L$ group. More than one allows the
possibility
 of distinguishing the global symmetry groups of the quarks and
leptons,
 or even different generations, naturally yielding a hierarchy of
mass scales.
 The point is that the various ETC scales are determined by the
chiral
 symmetry breaking scale at which the gauge and global symmetries
 combine. These need not be the same for all the generators.
 In CTSM--like models,

one would expect at the
 very least that there is a global
 symmetry group acting on the  two light families of quarks. However,
 it could be consistent to separate the third generation
\cite{randall}.
 Indeed, if there are different scales associated with  leptons and
quarks,
 or quarks of different generations, experimental bounds would have
 to be reevaluated. For example, compositeness bounds could be
 reinterpreted in terms of bounds on the ETC scale. But it is
 possible that the lepton scale is higher than the quark scale,
 so that stringent bounds on lepton compositeness are not a problem.
 Also, modifications of $Z$ couplings to light fermions would not
 all be determined at the same scale, weakening the stringency
 of the bounds of ref. \cite{mesek}.

  \section{Conclusion}
We have shown by explicit construction that one can have an ETC model
with a GIM mechanism. Furthermore, the first model is quite simple,
and the complications of the second class of models seem
necessary if one is to have high energy flavor violation
which can feed to low energy through mass terms, but include
a nontrivial KM matrix. We have also shown how to maintain
massless neutrinos and preserve a custodial SU(2) symmetry.
It therefore seems worthwhile to investigate whether this program
can be carried further. Many interesting questions remain,
such as how to generate the flavor physics, how the vacuum
alignment works,
and why the top quark is so heavy.

Our models might be criticized on the grounds of the additional
particles required. It should be recognized that this is an
inevitable consequence of constructing a model which is sufficiently
simple to analyze, since this would require well separated physics
scales with fermions to communicate between them.  It is
also important to recognize that none of the additional particles
transform under electroweak SU(2).  The flavor physics has
been completely isolated from the SU(2) breaking sector.

We conclude that this could be a promising direction for future
investigations
of technicolor.

The models very simply incorporate those facts which have
been proven true of the generalized Higgs sector. These ideas should
be
incorporated into future theoretical and
experimental investigations of technicolor.

  \section*{Acknowledgements}
  This work benefited greatly from  criticisms

  and suggestions from
  Howard Georgi. I am also grateful to David Kaplan,

Ann Nelson, and Mike Dugan

for useful comments. I  thank Mike Dugan and Nuria Rius for
comments on the manuscript. I thank the Aspen Center for Physics
and the University of Santa Cruz for their hospitality while this
work was being completed.
  This work is supported in part by funds provided by the U. S.
Department
  of Energy (D.O.E.) under contract \#DE--AC02--76ERO3069.

\def\thefiglist#1{\section*{Figure Captions\markboth
 {FIGURE CAPTIONS}{FIGURE CAPTIONS}}\list
 {Figure \arabic{enumi}.}
 {\settowidth\labelwidth{Figure #1.}\leftmargin\labelwidth
 \advance\leftmargin\labelsep
 \usecounter{enumi}\parsep 0pt \itemsep 0pt plus2pt}
 \def\newblock{\hskip .11em plus .33em minus -.07em}
 \sloppy}
\let\endthefiglist=\endlist

\begin{thefiglist}{9}
\item Moose notation: Circle is gauged SU(N) and two line segments
are fundamental and antifundamental representations of SU(N)
\item Models 1 and 2
\item Model 2 below $\mu_n$
\item Modified Model 2 with heavier pseudogoldstone bosons
\end{thefiglist}

 \end{document}